\documentclass[11pt]{article}
\usepackage{axodraw}
\usepackage{bbm} 
\usepackage{amsmath} 

\newcommand{\Tr}{{\rm Tr\,}}
\newcommand{\Lcal}{{\cal L}}
\def\lsi{\raise0.3ex\hbox{$<$\kern-0.75em\raise-1.1ex\hbox{$\sim$}}}
\def\gsi{\raise0.3ex\hbox{$>$\kern-0.75em\raise-1.1ex\hbox{$\sim$}}}
\newcommand{\lsim}{\mathop{\lsi}}

\newcommand{\la}[1]{\label{#1}}
\newcommand{\be}{\begin{equation}}
\newcommand{\ee}{\end{equation}}
\newcommand{\ba}{\begin{eqnarray}}
\newcommand{\ea}{\end{eqnarray}}
\newcommand{\fig}{Fig.~}
\newcommand{\eq}{Eq.~}

\newcommand{\app}{App.~}
\newcommand{\eqs}{Eqs.~}
\newcommand{\nr}[1]{(\ref{#1})}
\newcommand{\nn}{\nonumber \\}
\renewcommand{\(}{\left(}
\renewcommand{\)}{\right)}
\newcommand{\lb}{\left\{}
\newcommand{\rb}{\right\}}
\newcommand{\lk}{\left[}
\newcommand{\rk}{\right]}
\newcommand{\ld}{\left.}
\newcommand{\rd}{\right.}
\newcommand{\e}{\epsilon}
\newcommand{\order}[1]{{\cal O}(#1)}
\newcommand{\dilog}{{\rm Li}_2}
\newcommand{\approxDDD}{{\stackrel{d=3\!-\!2\e}{\approx}}}
\newcommand{\rmi}[1]{{\mbox{\scriptsize #1}}}
\newcommand{\PiT}{\Pi_\rmi{T}}

\newcommand{\picmm}[1]{\;\parbox[c]{35pt}{\begin{picture}(35,10)(0,0)
\SetWidth{1.0}\SetScale{1.0} #1 \end{picture}}\;}
\newcommand{\picms}[1]{\;\parbox[c]{35pt}{\begin{picture}(35,20)(0,0)
\SetWidth{1.0}\SetScale{1.0} #1 \end{picture}}\;}
\newcommand{\picmsa}[1]{\;\parbox[c]{35pt}{\begin{picture}(35,25)(0,0)
\SetWidth{1.0}\SetScale{1.0} #1 \end{picture}}\;}
\newcommand{\pics}[1]{\;\parbox[c]{20pt}{\begin{picture}(20,20)(0,0)
\SetWidth{1.0}\SetScale{1.0} #1 \end{picture}}\;}
\newcommand{\picssb}[1]{\;\parbox[c]{35pt}{\begin{picture}(35,35)(0,0)
\SetWidth{1.0}\SetScale{1.0} #1 \end{picture}}\;}
\newcommand{\pisca}[1]{\;\parbox[c]{27.5pt}{\begin{picture}(30,25)(0,0)
\SetWidth{1.0}\SetScale{1.0} #1 \end{picture}}\;}
\newcommand{\pisc}[1]{\;\parbox[c]{30pt}{\begin{picture}(30,20)(0,0)
\SetWidth{1.0}\SetScale{1.0} #1 \end{picture}}\;}
\newcommand{\pibcb}[1]{\;\parbox[c]{45pt}{\begin{picture}(45,20)(0,0)
\SetWidth{1.0}\SetScale{1.0} #1 \end{picture}}\;}

\newcommand{\picbb}[1]{\;\parbox[c]{35pt}{\begin{picture}(20,35)(0,0)
\SetWidth{1.0}\SetScale{1.0} #1 \end{picture}}\;}

\def\Lwidth{1}

\def\Agl(#1,#2)(#3,#4,#5){\PhotonArc(#1,#2)(#3,#4,#5){\Lwidth}
{6.283 #3 mul 360 div #4 #5 sub #4 #5 sub mul sqrt mul Ldensity mul}}
\def\Lgl(#1,#2)(#3,#4){\Photon(#1,#2)(#3,#4){\Lwidth}
{#1 #3 sub #1 #3 sub mul #2 #4 sub #2 #4 sub mul add sqrt Ldensity mul}}
\def\Agh(#1,#2)(#3,#4,#5){\DashArrowArc(#1,#2)(#3,#4,#5){1}}
\def\Ahg(#1,#2)(#3,#4,#5){\DashArrowArcn(#1,#2)(#3,#5,#4){1}}
\def\Lgh(#1,#2)(#3,#4){\DashArrowLine(#1,#2)(#3,#4){1}}
\def\Lhg(#1,#2)(#3,#4){\DashArrowLine(#3,#4)(#1,#2){1}}
\def\Ahh(#1,#2)(#3,#4,#5){\DashCArc(#1,#2)(#3,#4,#5){1}}
\def\Lhh(#1,#2)(#3,#4){\DashLine(#1,#2)(#3,#4){1}}
\def\Asc(#1,#2)(#3,#4,#5){\CArc(#1,#2)(#3,#4,#5)}
\def\Lsc(#1,#2)(#3,#4){\Line(#1,#2)(#3,#4)}

\def\Textt(#1,#2,#3){\Text(#1,#2)[t]{{$\scriptstyle #3$}}}
\def\Textb(#1,#2,#3){\Text(#1,#2)[b]{{$\scriptstyle #3$}}}

\def\ToptVS(#1,#2,#3){\pics{#1(10,10)(10,0,180) #2(10,10)(10,180,360)%
 #3(20,10)(0,10)}}
\def\TopoS(#1){\picmm{#1(0,5)(15,5) #1(20,5)(35,5)%
 \GCirc(17.5,5){3}{0}}}
\def\TopoSB(#1,#2,#3){\picms{#1(0,10)(7.5,10) #2(17.5,10)(10,0,180)%
 #3(17.5,10)(10,180,360) #1(27.5,10)(35,10)}}
\def\TopoSBo(#1,#2,#3){\picms{#1(0,10)(7.5,10) #2(17.5,10)(10,0,180)%
 #3(17.5,10)(10,180,360) #1(27.5,10)(35,10) \GCirc(17.5,20){3}{0}}}
\def\TopoSBu(#1,#2,#3){\picms{#1(0,10)(7.5,10) #2(17.5,10)(10,0,180)%
 #3(17.5,10)(10,180,360) #1(27.5,10)(35,10) \GCirc(17.5,0){3}{0}}}
\def\TopoSBl(#1,#2,#3){\picms{#1(0,10)(7.5,10) #2(17.5,10)(10,0,180)%
 #3(17.5,10)(10,180,360) #1(27.5,10)(35,10) \GCirc(27.5,10){3}{0}}}
\def\TopoSBr(#1,#2,#3){\picms{#1(0,10)(7.5,10) #2(17.5,10)(10,0,180)%
 #3(17.5,10)(10,180,360) #1(27.5,10)(35,10) \GCirc(7.5,10){3}{0}}}
\def\TopoSBlabel(#1,#2,#3,#4,#5){\picms{#1(0,10)(7.5,10) #2(17.5,10)(10,0,180)%
 #3(17.5,10)(10,180,360) #1(27.5,10)(35,10)%
 \Textt(17.5,18,#4) \Textb(18.5,2,#5)}}
\def\TopoST(#1,#2){\pisc{#1(0,0)(15,0) #1(15,0)(30,0)%
 #2(15,10)(10,-90,270)}} 
\def\TopoSTo(#1,#2){\pisc{#1(0,0)(15,0) #1(15,0)(30,0)%
 #2(15,10)(10,-90,270) \GCirc(15,20){3}{0}}} 
\def\TopoSTlabel(#1,#2,#3){\pisc{#1(0,0)(15,0) #1(15,0)(30,0)%
 #2(15,10)(10,-90,270) \Textt(15,17,#3)}} 
\def\ToptSM(#1,#2,#3,#4,#5,#6){\picms{#1(0,10)(7.5,10) #1(27.5,10)(35,10)%
 #2(17.5,10)(10,0,90) #3(17.5,10)(10,90,180) #4(17.5,10)(10,180,270)%
 #5(17.5,10)(10,270,360) #6(17.5,20)(17.5,0)}}
\def\ToptSAl(#1,#2,#3,#4,#5){\picms{#1(0,10)(7.5,10) #1(27.5,10)(35,10)%
 #2(17.5,10)(10,0,90) #3(17.5,10)(10,90,180) #4(17.5,10)(10,180,360)%
 #5(7.5,20)(10,270,360)}}
\def\ToptSAr(#1,#2,#3,#4,#5){\picms{#1(0,10)(7.5,10) #1(27.5,10)(35,10)%
 #2(17.5,10)(10,0,90) #3(17.5,10)(10,90,180) #4(17.5,10)(10,180,360)%
 #5(27.5,20)(10,180,270)}}
\def\ToptSE(#1,#2,#3,#4,#5){\pibcb{#1(0,10)(7.5,10) #1(37.5,10)(45,10)%
 #3(15,10)(7.5,0,180) #4(15,10)(7.5,180,360)%
 #2(30,10)(7.5,0,180) #5(30,10)(7.5,180,360)}} 
\def\ToptSS(#1,#2,#3,#4){\picms{#1(0,10)(7.5,10) #1(27.5,10)(35,10)%
 #4(7.5,10)(27.5,10) #2(17.5,10)(10,0,180) #3(17.5,10)(10,180,360)}}
\def\ToptSBB(#1,#2,#3,#4,#5,#6){\picmsa{#1(0,10)(7.5,10)%
 #2(17.5,10)(10,0,45) #3(17.5,10)(10,135,180) #4(17.5,10)(10,180,360)%
 #6(17.5,17.5)(7.5,0,180) #5(17.5,17.5)(7.5,180,360) #1(27.5,10)(35,10)}}
\def\ToptSBT(#1,#2,#3,#4,#5){\picssb{#1(0,10)(7.5,10)%
 #2(17.5,10)(10,0,90) #2(17.5,10)(10,90,180) #4(17.5,10)(10,180,360)%
 #5(17.5,27.5)(7.5,-90,270) #1(27.5,10)(35,10)}}
\def\ToptSTB(#1,#2,#3,#4,#5){\pisca{#1(0,0)(15,0) #1(15,0)(30,0)%
 #2(15,10)(10,-90,45) #3(15,10)(10,135,270)%
 #5(15,17.5)(7.5,0,180) #4(15,17.5)(7.5,180,360)}} 
\def\ToptSTT(#1,#2,#3,#4){\picbb{#1(0,0)(15,0) #1(15,0)(30,0)%
 #2(15,10)(10,-90,90) #3(15,10)(10,90,-90) #4(15,27.5)(7.5,-90,270)}} 
\def\ToptSDBl(#1,#2,#3,#4){\picms{#1(0,10)(7.5,10) #1(27.5,10)(35,10)%
 #2(17.5,10)(10,0,180) #3(17.5,10)(10,180,360) #4(12.5,10)(5,-90,270)}}
\def\ToptSDBr(#1,#2,#3,#4){\picms{#1(0,10)(7.5,10) #1(27.5,10)(35,10)%
 #2(17.5,10)(10,0,180) #3(17.5,10)(10,180,360) #4(22.5,10)(5,-90,270)}}

\begin{document}
 
\begin{titlepage}
\begin{flushright}
BI-TP-2012-12\\March 29, 2012
\end{flushright}
\begin{centering}
\vfill
                   
{\bf\Large Resummation scheme for 3d Yang-Mills and the 
two-loop magnetic mass for hot gauge theories}

\vspace{0.8cm}

Daniel Bieletzki$^{1}$, Kilian Lessmeier$^2$, Owe Philipsen$^1$, York Schr\"oder$^2$

\vspace{0.3cm}
{\em $^{\rm 1}$
Institut f\"ur Theoretische Physik, Goethe-Universit\"at Frankfurt,\\
Max-von-Laue-Str.~1, 60438 Frankfurt am Main, Germany}\\
{\em $^{\rm 2}$
Fakult\"at f\"ur Physik, Universit\"at Bielefeld, D-33501 Bielefeld, Germany}

\vspace*{0.7cm}
 
\begin{abstract}
Perturbation theory for non-Abelian gauge theories at finite temperature
is plagued by infrared divergences caused by magnetic soft modes 
$\sim g^2T$, which correspond to the fields of a 3d Yang-Mills theory. 
We revisit a gauge invariant resummation scheme 
to solve this problem by self-consistent mass generation using an auxiliary scalar field, 
improving over previous attempts in two respects. First, we generalise earlier
SU($2$) treatments to SU($N$).
Second, we obtain a gauge independent two-loop gap equation, correcting
an error in the literature. The resulting two-loop 
approximation to the magnetic mass represents a $\sim 15\%$ correction to the
leading one-loop value, indicating a reasonable convergence of the resummation.
\end{abstract}
\end{centering}
 
\vfill
 

\vfill

\end{titlepage}
 

\section{Introduction}

From heavy ion collision experiments to astroparticle physics and cosmology, there 
is a need for theoretical predictions from finite temperature gauge theories 
within and beyond the standard model. Since Monte Carlo
simulations of lattice gauge theories do not work for finite baryon densities or dynamical problems 
involving real time,  analytical approaches are warranted as well.
Perturbation theory for equilibrium quantities of non-Abelian gauge theories at finite temperature $T$ features three scales: a hard scale $\pi T$ of the non-zero Matsubara modes 
related to the compactified 
Euclidean time direction, the soft and ultra-soft scales $gT, g^2T$, with gauge coupling $g$, which are associated with the screening 
of colour-electric and colour-magnetic gauge fields, respectively. Since the dimensionless expansion parameter for the latter features a mass (or momentum) scale in the denominator, $\sim g^2T/E(\bf{p})$,  
prohibitive infrared divergences occur in the  
perturbative series when bare, massless propagators are used. 
This is known as the Linde problem \cite{linde, py}. 
Since  the dynamically generated magnetic screening mass is itself 
$m\sim g^2T$,
the problem cannot be cured to any finite order in ordinary perturbation theory. On the other hand,
evidence from gauge fixed lattice simulations 
(see \cite{teresa} and references therein) as well as gauge invariant simulations
of field strength correlators \cite{op01} 
is consistent with an effectively massive gluon propagator. 
Moreover, a dynamically generated gauge boson mass plays a prominent role
in a Hamiltonian description of three-dimensional Yang-Mills theory \cite{kara,Nair:2012zb} and is of
renewed interest in the context of a Higgs-less electroweak gauge sector, see e.g.~\cite{Kondo:2012wf}.
 
In this work we systematise a resummation method for 3d 
Yang-Mills theory, viz.~the magnetic sector of finite $T$ gauge theories, which has been 
proposed some time ago \cite{bp}. The general idea is to screen the infrared divergences by  
adding a gauge invariant mass term, which gets subtracted again at higher order, resulting in a resummation of the 
loop expansion. The mass term is not unique and several possibilities have been tried at one-loop
\cite{an,jp,corn}.
Under the name of screened perturbation theory similar techniques have been 
applied to scalar theories \cite{karsch,andersen} and to the colour-electric sector $\sim gT$ of gauge theories in a dimensionally reduced setting \cite{blaizot} as well as in four dimensions \cite{ja2}. 
There, a screening mass is generated in ordinary perturbation theory
and the resummation of its corrections is merely used as a means to improve the 
convergence. By contrast, the magnetic mass $\sim g^2T$ is entirely non-perturbative. It has to 
be generated by an infinite resummation and evaluated self-consistently by
a gap equation. Since the gauge coupling drops out of the 
effective expansion parameter, 
there is no parameter to tune and hence no regime where the resummation is guaranteed to 
work. Its convergence properties can be judged only after explicit calculations.

In this paper we present a systematic derivation of a gauge invariant resummation 
scheme using auxiliary fields based on the non-linear sigma model. In particular, we discuss the BRS
invariance of the resummation scheme which ensures that the symmetries of the theory are maintained
throughout a modified perturbative treatment.
We compute the gap equation for the magnetic screening mass through two loops in general $R_\xi$
gauges and duly find it to be gauge parameter independent, thus correcting an error in an earlier
two-loop investigation \cite{eber}. The two-loop result amounts to a $\sim 15$\%  correction to the 
leading one-loop result, thus pointing at a reasonable convergence of the resummation scheme.


\section{Resummation based on a non-linear sigma model}

Let us now focus on Euclidean Yang-Mills theory,
\be
\Lcal(A)=-\frac{1}{2g^2}\; \Tr F_{\mu\nu}^2\;,  
\ee
where we use a matrix notation, 
\be
F_{\mu\nu}=[D_\mu,D_\nu]\;,\quad 
D_\mu =\mathbbm{1}\partial_\mu-A_\mu(x)\;,\quad 
A_\mu(x)=igT^aA_\mu^a(x)\;,
\ee
with hermitean SU($N$) generators $T^a$, $a=1\dots N^2\!-\!1$.
We are interested in three dimensions, $\mu,\nu=1\ldots3$, where the coupling constant $g^2$
carries dimension of mass and is to be identified with $g^2=g_{\rm 4d}^2T$, if the action is viewed as
the magnetic part of hot Yang-Mills theory.
 
The general idea of a resummation is to sum up higher order contributions (infinitely many in our case) into a given order of a perturbative expansion. In order to avoid double counting,
these contributions must then be left out at the higher order where they naturally occur, such that the 
perturbative scheme gets reorganised in some systematic way. (For a discussion of various
schemes used in the context of thermal field theory, see \cite{bir}). 
This can be formalised by rewriting the Lagrangian serving for the perturbative expansion as \cite{jp} 
\be
\Lcal_{\rm eff}=\frac{1}{\ell}\lk\Lcal(\sqrt{\ell}X)+\Delta\Lcal(\sqrt{\ell}X)-\ell\Delta\Lcal(\sqrt{\ell}X)\rk\;,
\la{eq:resum}
\ee
where $X$ generically stands for the fields. The modification
$\Delta\Lcal$ contains fields of the original Lagrangian and possibly auxiliary fields.
In particular, if $\Delta\Lcal$
is chosen to represent a mass term for the gluon, this will regulate the infrared divergences.
The counting parameter $\ell$ in which one expands is to be set to $\ell=1$ at the end of a
calculation, for which the Lagrangian is identical to the original one. 
However, in a perturbative evaluation to finite order the results will
differ from the unresummed ones,  
the original theory being recovered exactly only at asymptotically high orders. 
Whether low order calculations are a good 
approximation to the full answer has to be judged from the apparent convergence of the 
series and may depend on the observable, just as in 
ordinary perturbation theory. 

A valid resummation scheme
has to maintain the symmetries of our gauge theory at every order, 
for general $\ell$. Clearly, this leaves many 
conceivable choices for $\Delta\Lcal$, several of which have been tried at one-loop level 
in the literature \cite{bp,an,jp,corn}.
An optimal choice would be based on convergence properties in higher orders. 
Here we work with a gauged non-linear sigma model,
coupling a field $\Phi\in {\rm U}(N)$ as
\be
\la{eq:defPi}
\Delta\Lcal(A,\pi)=\frac{m^2}{g^2}\Tr [(D_\mu\Phi)^\dag D_\mu\Phi]\;,\quad 
\Phi^\dagger\Phi=\mathbbm{1}\;,\quad
\Phi(x)=e^{\pi(x)}\;,\quad
\pi(x)=i\frac{g}{m}T^a\pi^a(x)\;,
\ee 
where the $\pi^a(x)\in\mathbbm{R}$ are auxiliary would-be Goldstone boson
fields\footnote{Note that it is the $\pi$ and not the $\Phi$ who get rescaled as $\pi\rightarrow \sqrt{\ell}\pi$ for the purpose of resummation, \eq\nr{eq:resum}.} with the same mass dimension
as the gauge fields $A^a_\mu(x)$.  
Under gauge transformations, 
$\Phi'=U\Phi$ and $(D_\mu\Phi)'=U(D_\mu\Phi)$, with the unitary matrix $U=e^{\Lambda(x)}$ and corresponding real coefficients $\Lambda^a (x)$, with $\Lambda(x)=iT^a\Lambda^a (x)$. Thus,
$\Delta\Lcal$ provides a mass term for the gauge fields at tree-level while maintaining gauge invariance.


\section{Gauge fixing and BRS-invariance}
\la{se:gfbrs}

In order to do perturbative calculations, a gauge needs to be fixed. 
It is well known that in Higgs and sigma models the standard covariant gauges lead to non-diagonal or mixing terms 
in scalar and gauge fields, $\sim (\partial_\mu \pi) A_\mu$. This can be avoided by
choosing $R_\xi$-gauges.   
In the case of a resummed calculation,
however, additional choices have to be made. 

We can either take the point of view that our starting point is
the resummed theory before gauge fixing as in \eq\nr{eq:resum}, and then add gauge fixing and ghost terms to that expression,
\ba
 \Lcal_{\rm eff,A}&=&\frac{1}{\ell}\lb\Lcal(\sqrt{\ell}A)+(1-\ell)\lk\Delta\Lcal(\sqrt{\ell}A,\sqrt{\ell}\pi) 
 \rd\rd\nn&&\;\hspace{0.5cm}\ld\ld
 +\Lcal_{\rm gf,A}(\sqrt{\ell}A,\sqrt{\ell}\pi)+\Lcal_{\rm FP,A}(\sqrt{\ell}A,\sqrt{\ell}\pi,\sqrt{\ell}c)\rk\rb\;,\\
\la{eq:gfA}
\Lcal_{\rm gf,A}(A,\pi)&=&
 -\frac1{g^2\xi}\Tr\lb\((\partial_\mu A_\mu)
 -\xi m^2 \Tr\((\Phi-\Phi^{\dagger})T^a\)
 T^a\)^2\rb\;,\\
\Lcal_{\rm FP,A}(A,\pi,c)&=&
 \frac{1}{g^2}\Tr\lb2(\partial_\mu\bar{c})\((\partial_\mu c)
 -[A_\mu,c]\)+ \xi m^2\bar{c}\(\Phi^{\dagger}c
 +c\Phi\)\rb\;,
\ea
with ghost fields $c=igT^ac^a$ and anti-ghost fields $\bar{c}=-igT^a\bar{c}^a$.
We refer to this gauge fixing procedure as A. For $\ell=1$ we obtain Yang-Mills theory without
gauge fixing, as in \eq\nr{eq:resum}. This invokes the following Feynman rules for the 
counter term two-point vertices,
\ba
\la{eq:ct1}
\TopoS(\Lgl)&:&\quad\Gamma^{ab}_{\mu\nu,{\rm A}}(A^2)\;=\;
\(\frac{p_\mu p_\nu}{\xi}
+m^2\delta_{\mu\nu}\) \ell\delta^{ab}\;,\\
\la{eq:ct2}
\TopoS(\Lsc)&:&\quad\Gamma^{ab}_{\rm A}(\pi^2)\;=\;(p^2+\xi m^2) \ell\delta^{ab}\;,\\
\la{eq:ct3}
\TopoS(\Lgh)&:&\quad\Gamma^{ab}_{\rm A}(c^2)\;=\;(p^2+\xi m^2) \ell\delta^{ab}\;.
\ea

Alternatively, we may consider 
Yang-Mills theory in a covariant gauge and then resum the gauge fixed theory,
a strategy which we refer to as procedure B, 
\ba
\la{eq:action}
 \Lcal_{\rm eff,B}&=&\frac{1}{\ell}\lb\Lcal(\sqrt{\ell}A)+(1-\ell)\Delta\Lcal(\sqrt{\ell}A,\sqrt{\ell}\pi) \rd
\nn&&\;\hspace{0.5cm}\ld
 +\Lcal_{\rm gf,B}(\sqrt{\ell}A,\sqrt{\ell}\pi)+\Lcal_{\rm FP,B}(\sqrt{\ell}A,\sqrt{\ell}\pi,\sqrt{\ell}c)\rb\;,\\
\la{eq:gfB}
\Lcal_{\rm gf,B}(A,\pi)&=&
 -\frac1{g^2\xi} \Tr\lb\((\partial_\mu A_\mu)
 -(1-\ell)\xi m^2 \Tr\((\Phi-\Phi^{\dagger})T^a\)
 T^a\)^2\rb\;,\\
\Lcal_{\rm FP,B}(A,\pi,c)&=&
 \frac{1}{g^2}\Tr\lb2(\partial_\mu\bar{c})\((\partial_\mu c)
 -[A_\mu,c]\) +(1-\ell)\xi m^2 
 \bar{c}\(\Phi^{\dagger}c+c\Phi\) \rb\;.
\ea
The corresponding counter term two-point vertices read
\ba
\la{eq:newct1}
\TopoS(\Lgl)&:&\quad\Gamma^{ab}_{\mu\nu,{\rm B}}(A^2)\;=\;
(m^2\delta_{\mu\nu}) \ell\delta^{ab}\;,\\
\la{eq:newct2}
\TopoS(\Lsc)&:&\quad\Gamma^{ab}_{\rm B}(\pi^2)\;=\;
(p^2+\xi (2-\ell)m^2) \ell\delta^{ab}\;,\\
\la{eq:newct3}
\TopoS(\Lgh)&:&\quad\Gamma^{ab}_{\rm B}(c^2)\;=\;
(\xi m^2) \ell\delta^{ab}\;.
\ea

The two formulations feature non-trivial differences.
Note that \eq\nr{eq:newct2} 
contributes to both, order $\sim \ell^1$ and  $\sim\ell^2$.
Let us remark here that the gauge fixing and corresponding counter terms 
used in \cite{bp,eber} work only for
the gluon pole mass to leading and next-to-leading order, but 
require generalisation (as above) 
for higher orders and other observables.

In order to provide gauge invariant results for physical observables order by order in perturbation theory, 
it is necessary and sufficient that the gauge fixed Lagrangian $\Lcal_{\rm eff}$ is invariant under 
BRS-transformations \cite{brs,zinnj}. 
 Note that in the resummed theory with general $\ell$ all fields get rescaled
 by $\sqrt{\ell}$, and so do the gauge transformations.  A BRS-transformation now corresponds to
 the special choice $\Lambda=\omega\sqrt{\ell} c$. The variations of
 fields and Faddeev-Popov ghosts under infinitesimal BRS-transformations
 are
\ba
\la{eq:BRSsun}
\delta_{\rm B} A_\mu &=& \omega(\partial_\mu c)
+\omega\sqrt{\ell}\(cA_\mu-A_\mu c\)\;,\nn
\delta_{\rm B} \Phi &=& \omega\sqrt{\ell} c \Phi \nn
\Rightarrow&&
\delta_{\rm B} \pi \;=\; \omega\sum_{n=0}^\infty 
\frac{B_n\,\ell^{\frac{n}2}}{n!}
\sum_{j=0}^n (-1)^j {n\choose j}\pi^{n-j} c \pi^j
+\order{\omega^2}\nn&&\hphantom{\delta_{\rm B} \pi }
\;\approx\; \omega c
-\frac{\omega\sqrt{\ell}}2\(\pi c-c\pi\)
+\frac{\omega\ell}{12}\(\pi\pi c-2\pi c\pi+c\pi\pi\)
+\order{\pi^4,\omega^2}\;,\nn
\delta_{\rm B} c &=& \omega\sqrt{\ell} c c\;,\nn
\delta_{\rm B} \bar c &=& -\frac\omega{\xi}\((\partial_\mu A_\mu)
-(1-\ell)\xi m^2\Tr\((e^{\sqrt{\ell}\pi}-e^{-\sqrt{\ell}\pi})T^a\)
T^a/\sqrt{\ell}\)\;,
\ea
where $B_n$ are the Bernoulli numbers.
The above transformation refers to $\Lcal_{\rm eff,B}$; for formulation A,
the factor $(1-\ell)$ in $\delta_{\rm B} \bar c$ is absent.

As a non-trivial check we have performed our calculations in both ways, obtaining identical gauge invariant results for both setups. We present our
results according to setup B, as it is closer to the standard perturbative treatment.


\section{Relation to SU($2$) calculations}

Here, we connect our general SU($N$) parametrisation of the scalar field,
\eq\nr{eq:defPi},
to the special case of SU($2$) treated in \cite{bp,eber}.
Using $\Tr(T^a)=0$, $\Tr(T^aT^b)=\frac12\delta^{ab}$ and 
$\Tr\mathbbm{1}=N$, one gets for the product of two generators 
the standard expression
\ba
T^a T^b &=& 
\frac{\delta^{ab}}{2N}\,\mathbbm{1} 
+\Tr(\{T^a,T^b\}T^c)\,T^c 
+\Tr([T^a,T^b]T^c)\,T^c \nn
&\equiv& \frac{\delta^{ab}}{2N}\,\mathbbm{1} 
+\frac12\,d^{abc}\,T^c\hphantom{TTTTTT}+\frac12\,i\,f^{abc}\,T^c \;.
\ea
Note that for the special case of SU($2$), where $T^a\sim\sigma^a$
and the Pauli matrices anticommute as 
$\{\sigma^a,\sigma^b\}=2\delta^{ab}\mathbbm{1}$, 
the totally symmetric structure constants vanish, $d^{abc}=0$.
Hence, in SU($2$) the product of two of our scalar fields is diagonal,
\ba
\pi\,\pi 
= (ig/m)^2 \pi^a\pi^b\,\frac12\(\frac{\delta^{ab}}{N}\,\mathbbm{1}
+d^{abc}\,T^c\)
&\stackrel{{\rm SU}(2)}=& (ig/2m)^2 \pi^a\pi^a\,\mathbbm{1} \;,
\ea
and therefore the field $\Phi$ can be expressed as
\ba
\Phi &=& e^\pi 
\;=\; \cos\frac\pi{i}+i\sin\frac\pi{i}
\;\;\stackrel{{\rm SU}(2)}=\; \;\sigma\,\mathbbm{1}+i\,\bar{\pi}^{a}\,T^a\\
&&{\rm with}\quad\sigma \equiv \cos\frac{g\sqrt{\pi^a\pi^a}}{2m} 
\approx 1+\order{\pi^2}\;,\\
&&{\rm and}\quad
\bar{\pi}^{a} \equiv \frac{2\pi^a}{\sqrt{\pi^a\pi^a}}
\sin\frac{g\sqrt{\pi^a\pi^a}}{2m} \approx \frac{g}m\,\pi^a
+\order{\pi^3}\;.
\ea
Hence, for $N=2$ our model can be recast into the form of
the model considered in \cite{bp}. For general $N$, however,
the coefficients on the right-hand side of 
$\Phi=\frac{\Tr\Phi}N\,\mathbbm{1}+2\Tr(\Phi T^a)\,T^a$
can not be expressed in a closed form in terms of the real fields 
$\pi^a$.
The traceless part of $\Phi$ is part of our construction
for the gauge fixing term, see \eqs\nr{eq:gfA} and \nr{eq:gfB},
which represents a non-trivial generalisation of 
$R_\xi$ gauges to SU($N$).


\section{Pole mass from a gauge invariant gap equation}

Having designed a general 
gauge invariant resummation scheme for 3d Yang-Mills theory, we now apply it to a calculation of
the gluon self-energy. 
Its transverse and longitudinal parts $\Pi_{\rm T/L}$ are defined as
\ba
\la{eq:Pimunu}
\Pi_{\mu\nu}^{ab}(p) &\equiv& \delta^{ab}
\lb\(\delta_{\mu\nu}-\frac{p_\mu p_\nu}{p^2}\)\PiT(p^2)
+\frac{p_\mu p_\nu}{p^2}\,\Pi_L(p^2)\rb\;.
\ea

The self-energy itself generally is a gauge dependent quantity.
However, the pole of the transverse
part of a gauge boson propagator,  $D_{\rm T}$, is known to be gauge invariant order by order in perturbation 
theory \cite{rebhan,jeger}, 
and we may employ our resummation scheme to evaluate it. (The longitudinal degrees of freedom
with a gauge dependent pole correspond to unphysical would-be Goldstone bosons as usual in Higgs-like theories, and can be gauged away in unitary gauge.) 
Without resummation the pole of the bare transverse propagator is at $p^2=0$,
whereas in the resummed theory it gets shifted to $p^2=-m^2$.
Identifying $m$ with the pole of the full 
propagator, we require that the pole
stays at $p^2=-m^2$ 
to any loop order. Taylor expanding the self-energy about the pole, the transverse propagator reads
\be
\la{eq:D_T}
D_{\rm T}=\frac1{p^2+m^2-\PiT(p^2)}
\stackrel{p^2=-m^2+\delta p^2}{=}
\frac{\frac1{1-\PiT'(-m^2)}}{-\frac{\PiT(-m^2)}{1-\PiT'(-m^2)}+\delta p^2+
\order{(\delta p^2)^2}} \;,
\ee
where $\PiT'(-m^2)\equiv \partial_{p^2} \PiT(p^2)|_{p^2=-m^2}$.
Near the pole it then corresponds to a massive propagator with residue
$Z(m^2)$,
\be
D_{\rm T}\propto\frac{Z(m^2)}{p^2+m^2}\;,\quad Z(m^2)=\frac1{1-\PiT'(-m^2)}\;,
\ee 
provided the first term in the denominator of \eq\nr{eq:D_T} vanishes.
This leads to the gap equation
\be
\la{eq:gap}
0\stackrel{!}{=}\frac{\PiT(-m^2)}{1-\PiT'(-m^2)}\;.
\ee 
Introducing the $\ell$-expansion of the self-energy, 
$\PiT=\sum_{n\ge1}\ell^n\,\PiT^{(n)}(p^2)$,
we now expand the gap equation to the desired order
 in $\ell$ and evaluate it after setting $\ell=1$.
Note that, to every order in $\ell$, the gap equation receives different kinds of self-energy contributions
contributing to order $\ell^n$.
In order to list these separately,
we introduce $\PiT^{(n-k),k}$ to denote the sum of diagrams with $(n-k)$ loops and $k$ counter term insertions,
\be
\la{eq:CTdef}
\PiT^{(n)}=\sum_{k=0}^{n}\PiT^{(n-k),k}\;.
\ee
The corresponding gap equation combines gauge dependent contributions of different self-energies 
$\PiT^{(n-k),k}$ into a gauge invariant quantity.

We perform our calculations using dimensional regularisation, working in $d$ dimensions and with generic gauge fixing parameter $\xi$. Details
and intermediate results are relegated to the appendix, from which we collect the results in the following 
sections.


\subsection{One-loop gap equation}

To leading order $\ell^1$, the gap equation \eq\nr{eq:gap} is simply
\be
\la{eq:gap1}
0=\PiT^{(1)}(-m^2)= 
\PiT^{(1),0}(-m^2)+\PiT^{(0),1}(-m^2)\;.
\ee
The five diagrams shown in \fig\ref{fig:1S0CT}
have been computed
in a general $R_\xi$-gauge in \cite{bp} and lead to a 
$\xi$-independent gap equation when evaluated on the pole.
From appendix \ref{se:details}, we reproduce these results for SU($N$) as
\ba
 \PiT^{(1),0}(-m^2) &\approxDDD& 
 \frac{g^2 N m}{8\pi}\(\frac34-\frac{63}{16}\ln3\) +\order{\e}\;,\\
\la{eq:pi01}
 \PiT^{(0),1}(p^2) &=& m^2\;.
\ea
Solving the quadratic one-loop gap equation 
\eq\nr{eq:gap1} then yields the well-known solutions
\ba
\la{eq:1loopSoln}
m_{\rm 1-loop}=0 \quad{\rm or}\quad
m_{\rm 1-loop}=\(\frac{63}{16}\ln3-\frac34\) \frac{g^2 N}{8\pi}
= 0.142276\,g^2 N \;.
\ea

\begin{figure}[t]
\begin{eqnarray*}
&&
 {1\over2}\TopoST(\Lgl,\Agl)
 +{1\over2}\TopoSB(\Lgl,\Agl,\Agl)
 -{1}\TopoSB(\Lgl,\Agh,\Ahh)
 +{1\over2}\TopoSB(\Lgl,\Asc,\Asc)
 \quad;\quad\TopoS(\Lgl)
\end{eqnarray*}
\caption{\la{fig:1S0CT} 
The self-energy diagrams contributing to order $\ell^1$.
Wiggly/dotted/full lines denote gluons/ghosts/scalars, respectively.
In the notation of \eq\nr{eq:CTdef}, the first four diagrams give 
$\PiT^{(1),0}$, while the last is $\PiT^{(0),1}$.}
\end{figure}


\subsection{Two-loop gap equation}

At order $\ell^2$, the gap equation  \eq\nr{eq:gap} reads
\ba
0&=&\PiT^{(1)}(-m^2)\(1+\partial_{p^2}\PiT^{(1)}(p^2)|_{p^2=-m^2}\)
+\PiT^{(2)}(-m^2)\\
&=&\(\PiT^{(0),1}+\PiT^{(1),0}\)\(1+\partial_{p^2}\PiT^{(1),0}
+\partial_{p^2}\PiT^{(0),1}\)+\PiT^{(2),0}+\PiT^{(1),1}\nn
&=& m^2 +\PiT^{(1),0} 
+\(\PiT^{(1),1}+m^2\partial_{p^2}\PiT^{(1),0}\)
+\(\PiT^{(2),0}+\PiT^{(1),0}\partial_{p^2}\PiT^{(1),0}\)
\;,\nonumber
\ea
where in the last line we have used \eq\nr{eq:pi01} and grouped together 
terms which will prove to be gauge invariant.
There are 38 genuine two-loop diagrams contributing to $\PiT^{(2),0}$, shown 
in \fig\ref{fig:2S0CT}. 
These can be expressed in terms of six scalar master integrals. 
Note that in unitary gauge ($\xi \rightarrow \infty$, to be taken before regularisation) the ghosts and 
pseudo-goldstones decouple leaving
only nine diagrams (see also \cite{eber}). 
\begin{figure}[t]
\begin{eqnarray*}
&&\hspace{-8mm}
{1\over4}\ToptSDBl(\Lgl,\Asc,\Asc,\Asc)
+{1\over4}\ToptSDBr(\Lgl,\Asc,\Asc,\Asc)
+{1\over4}\ToptSTT(\Lgl,\Agl,\Agl,\Agl)
+{1\over6}\ToptSS(\Lgl,\Agl,\Agl,\Lgl)
+{1\over6}\ToptSS(\Lgl,\Asc,\Asc,\Lsc)
+{1\over2}\ToptSAl(\Lgl,\Agl,\Agl,\Agl,\Agl)
+{1\over2}\ToptSAr(\Lgl,\Agl,\Agl,\Agl,\Agl)
\\[5mm]&&\hspace{-8mm}
+{1\over4}\ToptSTB(\Lgl,\Agl,\Agl,\Agl,\Agl)
+{1\over4}\ToptSTB(\Lgl,\Agl,\Agl,\Asc,\Asc)
-{1\over2}\ToptSTB(\Lgl,\Agl,\Agl,\Ahh,\Agh)
+{1\over4}\ToptSE(\Lgl,\Agl,\Agl,\Agl,\Agl)
+{1\over4}\ToptSE(\Lgl,\Asc,\Asc,\Asc,\Asc)
-{1\over2}\ToptSE(\Lgl,\Agh,\Asc,\Asc,\Ahh)
\\[5mm]&&\hspace{-8mm}
-{1\over2}\ToptSE(\Lgl,\Asc,\Agh,\Ahh,\Asc)
+{1\over2}\ToptSBT(\Lgl,\Agl,\Agl,\Agl,\Agl)
+{1\over2}\ToptSBT(\Lgl,\Asc,\Asc,\Asc,\Asc)
-{1}\ToptSBT(\Lgl,\Asc,\Asc,\Asc,\Agh)
-{1\over2}\ToptSBT(\Lgl,\Ahh,\Ahh,\Agh,\Asc)
-{1\over2}\ToptSBT(\Lgl,\Ahh,\Ahh,\Ahg,\Asc)
\\[5mm]&&\hspace{-8mm}
+{1\over2}\ToptSM(\Lgl,\Agl,\Agl,\Agl,\Agl,\Lgl)
+{1\over2}\ToptSM(\Lgl,\Asc,\Agl,\Agl,\Asc,\Lsc)
+{1\over2}\ToptSM(\Lgl,\Agl,\Asc,\Asc,\Agl,\Lsc)
+{1\over2}\ToptSM(\Lgl,\Asc,\Asc,\Asc,\Asc,\Lgl)
-{1}\ToptSM(\Lgl,\Ahh,\Agl,\Agl,\Ahh,\Lgh)
-{1}\ToptSM(\Lgl,\Agl,\Ahh,\Ahh,\Agl,\Lhg)
\\[5mm]&&\hspace{-8mm}
-{1}\ToptSM(\Lgl,\Agh,\Ahh,\Agh,\Ahh,\Lgl)
-{1}\ToptSM(\Lgl,\Ahh,\Asc,\Asc,\Ahh,\Lgh)
-{1}\ToptSM(\Lgl,\Asc,\Ahh,\Ahh,\Asc,\Lhg)
-{1}\ToptSM(\Lgl,\Agh,\Ahh,\Agh,\Ahh,\Lsc)
+{1\over2}\ToptSBB(\Lgl,\Agl,\Agl,\Agl,\Agl,\Agl)
+{1\over2}\ToptSBB(\Lgl,\Agl,\Agl,\Agl,\Asc,\Asc)
\\[5mm]&&\hspace{-8mm}
+{1}\ToptSBB(\Lgl,\Asc,\Asc,\Asc,\Agl,\Asc)
-{1}\ToptSBB(\Lgl,\Agl,\Agl,\Agl,\Ahh,\Agh)
-{1}\ToptSBB(\Lgl,\Ahh,\Ahh,\Agh,\Agl,\Ahh)
-{1}\ToptSBB(\Lgl,\Ahh,\Ahh,\Ahg,\Agl,\Ahh)
-{1}\ToptSBB(\Lgl,\Asc,\Asc,\Asc,\Ahh,\Agh)
-{1}\ToptSBB(\Lgl,\Ahh,\Ahh,\Agh,\Asc,\Ahh)
-{1}\ToptSBB(\Lgl,\Ahh,\Ahh,\Ahg,\Asc,\Ahh)
\end{eqnarray*}
\caption{\la{fig:2S0CT} 
The 38 diagrams contributing to $\PiT^{(2),0}$.
Notation as in \fig\ref{fig:1S0CT}.}
\end{figure}
The one-loop diagrams with one counter term insertion are shown in \fig\ref{fig:1S1CT}. 
A tree-level diagram with two counter term insertions is not one-particle-irreducible and hence does 
not contribute.
From appendix \ref{se:details} 
(where the renormalised 3d coupling $g^2(\mu)=\mu^{-2\e}g_{\rm bare}^2$ was introduced), 
the different contributions to the two-loop gap equation are
\ba
\la{eq:gap2}
\PiT^{(2),0}+\PiT^{(1),0}\partial_{p^2}\PiT^{(1),0}
&\approxDDD&
\(\frac{g^2N}{8\pi}\)^2\(\frac3{20\e}-10.6452
+\frac9{10}+\frac3{10}\ln\frac{\bar\mu^2}{4m^2}\)
+\order{\e}\;,\\
\PiT^{(1),1}&\approxDDD&
\frac{g^2 N m}{8\pi} \(\frac{21}{8}\ln3-\frac92+\frac{1-4\xi}{8}
\ln\frac{2\sqrt{\xi}+1}{2\sqrt{\xi}-1}+\frac32\sqrt{\xi}\)
+\order{\e}\;,\nn
\partial_{p^2}\PiT^{(1),0}&\approxDDD&
\frac{g^2 N}{8\pi m} \(-\frac{21}{32}\ln3+\frac{33}{8}-\frac{1-4\xi}{8}
\ln\frac{2\sqrt{\xi}+1}{2\sqrt{\xi}-1}-\frac32\sqrt{\xi}\)
+\order{\e}\;,
\nonumber
\ea
where all quantities are understood on-shell $(p^2=-m^2)$ and
for an analytic expression of the two-loop constant $10.6452$ 
we refer to \eq\nr{eq:2loopExp}.
Note that in general the on-shell self-energy is gauge dependent. However, the parts of the gap equation
pertaining to the non-linear sigma model without counter terms, i.e.~the first of \eqs(\ref{eq:gap2}), as well
as the sum of all counter term contributions are separately gauge invariant, thus leading to a gauge invariant solution for the pole mass in the
resummed theory. Our result for the second line of \eq(\ref{eq:gap2}) corrects an
error in an earlier calculation for $SU(2)$ \cite{eber}, which led to a
$\xi$-dependent pole.

\begin{figure}[t]
\begin{eqnarray*}
&&
 {1\over2}\TopoSTo(\Lgl,\Agl)
 +{1}\TopoSBo(\Lgl,\Agl,\Agl)
 -{1}\TopoSBo(\Lgl,\Agh,\Agh)
 -{1}\TopoSBu(\Lgl,\Agh,\Agh)
 +{1}\TopoSBo(\Lgl,\Asc,\Asc)
 +{1\over2}\TopoSBr(\Lgl,\Asc,\Asc)
 +{1\over2}\TopoSBl(\Lgl,\Asc,\Asc)
\end{eqnarray*}
\caption{\la{fig:1S1CT} 
The 7 diagrams contributing to $\PiT^{(1),1}$.
Notation as in \fig\ref{fig:1S0CT}.}
\end{figure}

The first of \eqs(\ref{eq:gap2}) features a divergence
as $d\rightarrow 3$, which is removed by mass renormalisation 
according to
$m_{\rm bare}^2=m^2(\mu)+\delta m^2$, where 
$\delta m^2=-\frac{3\,x^{2\e}}{20\e}\(\frac{g^2(\mu) N}{8\pi}\)^2$
with $x=1$ ($x=4\pi e^{-\gamma}$) 
specifying the MS ($\overline{\rm MS}$) scheme, respectively,
and $m^2(\mu)$ denotes the renormalised mass.
The renormalised two-loop gap equation reads (with $\bar\mu^2=4\pi e^{-\gamma}\mu^2$)
\ba
\la{eq:mass2} 
0&=&m^2+\frac{g^2 N m}{8\pi}\frac12\(\frac34-\frac{63}{16}\ln3\)
+\(\frac{g^2 N}{8\pi}\)^2\(-10.6452+\frac9{10}
+\frac3{10}\ln\frac{\bar\mu^2}{4\,x\,m^2}\)\;.
\ea
Note that the renormalisation prescription has introduced 
scheme as well as scale dependence.
In the following we pick the MS scheme ($x=1$).
The scale dependence is formally
of higher order: since the bare mass is
scale independent, the renormalised mass reacts to a scale variation as
$m^2(\mu)=m^2(\mu_0)-\frac35\(\frac{g^2 N}{8\pi}\)^2\ln(\mu/\mu_0)$. 
In a truncated perturbative series, however, this scale dependence
remains and can be taken as an estimate for the size of higher
order corrections. For a particular choice of scale, $\mu=m$,
the logarithm can be absorbed into the pole mass, 
such that the gap equation reduces to 
$0= m^2-0.07114 g^2N m-0.01516g^4N^2$, with positive solution
$m_{\rm 2-loop}=0.1637\,g^2 N$.

\eq\nr{eq:mass2} possesses non-trivial, real solutions which
are listed in Table \ref{tab:mass2S0CTres} 
(again in the MS scheme, $x=1$;
the second solution, $C_2$, is almost always close to zero, 
and we shall hence still call it trivial).
Since the gluon does not correspond to an asymptotic particle state, 
the scale dependence of its pole mass is expected and in complete
analogy to the two-loop pole masses of the electroweak gauge bosons \cite{jeger}. Note that the
pole mass changes by $\lsim 10\%$ only as the renormalisation scale is varied over two orders of 
magnitude. Together with the fact that the two-loop contribution constitutes a $\sim15\%$ correction
to the leading-order one-loop result, this points to a reasonable convergence of the resummation scheme.

\begin{table}[t]
\centering
  \begin{tabular}{| c | c | c | c | c | c |}\hline
    $\frac{\bar{\mu}}{g^2N}$ & $0.1$ &$0.3$ &$1$ &$3$ & $10$  \\ \hline
    $C_1$ &$0.1692$& $0.1651$& $0.1605$& $0.1562$& $0.1512$ \\ \hline
    $C_2$ &$4.4\cdot10^{-9}$& $1.3\cdot10^{-8}$& $4.4\cdot10^{-8}$&
    $1.3\cdot10^{-7}$& $4.4\cdot10^{-7}$ \\ \hline
 \end{tabular}
\caption{Scale-dependent solutions of the two-loop 
gap equation \eq\nr{eq:mass2}, $m=C_i\,g^2N$.}
\la{tab:mass2S0CTres}
\end{table}


\section{Conclusions}

We have generalised a non-perturbative resummation scheme for three-dimensional Yang-Mills theory
based on the non-linear sigma model \cite{bp} to SU($N$). 
Adding and subtracting a covariantly coupled 
scalar field allows for a gauge invariant gluon mass term regulating infrared divergences encountered
in bare perturbation theory.
We have established that this leads to gauge invariant
physical results by analysing the BRS-invariance of the resummed theory. 
As an application, we have calculated the
transverse gluon propagator and evaluated its pole mass by means of a gap equation, which we
explicitly verified to be gauge invariant through two loops, thus correcting an error in \cite{eber}. 
The pole mass requires normalisation. We have employed 
the minimal subtraction (MS) scheme, through which it 
acquires a  weak scale dependence. We find the two-loop correction to be $\sim 15\%$ of the leading
one-loop result, and the scale dependence $\sim 10\%$ of the two-loop result when the renormalisation
scale is varied over two orders of magnitude. Together, these two features might be indicative of a
reasonable convergence behaviour of our resummation scheme. 

As a further application, the scheme
lends itself to an evaluation of the $g^6$-contribution to the thermodynamic pressure in four-dimensional
gauge theories. Preliminary investigations up to two loops have been reported in \cite{Philipsen:2009dd},
a three-loop calculation is currently under way.


\section*{Acknowledgements}

D.B.~and O.P.~are supported by the German Bundesministerium f\"ur Bildung und Forschung (BMBF), grant no. 06MS9150.
Y.S.~is supported by the Heisenberg program of the Deutsche
Forschungsgemeinschaft (DFG), contract no.~SCHR~993/1. 


\begin{appendix} 


\section{Details of the calculation}
\la{se:details}
\newcommand{\pref}{\lk\frac{g_{\rm bare}^2\,N\,J(d,m)}{m^2(1-d)}\rk}

Due to the somewhat non-standard action with potentially high-order 
vertices, we automatically generate Feynman rules as well as a model
file directly from \eq\nr{eq:action}.
For two-loop self-energies, we potentially need vertices with up to six legs.

For notational simplicity, let us write the loop expansion 
of the bare on-shell transverse self-energy as defined in \eq\nr{eq:Pimunu} 
as well as its derivatives as
\ba
\la{eq:defhatPi}
\ld\partial_{p^2}^{\,a}\,\partial_{m^2}^{\,b}\,
\PiT^{\rm bare}(p^2)\right|_{p^2=-m^2} &=&
\(m^2\)^{1-a-b}
\sum_{n\ge 1}\pref^n\,
\hat\Pi_{ab}^{(n)}\;,
\ea
where the $\hat\Pi^{(n)}$ are dimensionless functions of $d, \xi$ and $N$ only 
which are computed from $n$\/-loop Feynman diagrams,
and $J(d,m)$ is the massive one-loop tadpole integral defined by
\eq\nr{eq:J}.
Note that the square bracket in dimensional regularisation expands as
\be
\pref \;\approxDDD\; 
\frac{g^2 N}{8\pi m}
\(\frac{\bar\mu}{2m}\)^{2\e}\(1+3\e+\order{\e^2}\)\;,
\ee
with renormalised coupling (note that $Z_{g^2}=1$ in 3d) 
$g^2=\mu^{-2\e}g_{\rm bare}^2$ 
and the usual $\overline{\rm MS}$ scale $\bar\mu^2=4\pi e^{-\gamma}\mu^2$.
From \eq\nr{eq:defhatPi} we explicitly have, 
in the notation of the main text\footnote{The last of the four relations
is non-trivial and follows from realising the mass-shift 
$m^2\rightarrow(1-\ell)m^2$ needed for the resummed theory \eq\nr{eq:action}
by a translation operator 
$\exp(p^2\ell\partial_{m^2})$ followed by the on-shell condition.},
\ba
\PiT^{(1),0}(-m^2) &=& 
m^2 \pref \hat\Pi_{00}^{(1)}\;,\\
\PiT^{(2),0}(-m^2) &=& 
m^2 \pref^2 \hat\Pi_{00}^{(2)}\;,\\
\ld\partial_{p^2}\PiT^{(1),0}(p^2)\right|_{p^2=-m^2} &=& 
\pref \hat\Pi_{10}^{(1)}\;,\\
\la{eq:Pi11CT}
\PiT^{(1),1}(-m^2) &=& 
-m^2 \pref \hat\Pi_{01}^{(1)}\;.
\ea

From here, the calculation proceeds via standard methods.
All diagrams that we need are generated with QGRAF \cite{qgraf}.
We then shift momenta to our conventions,
apply colour and Lorentz projectors, perform colour traces via 
the Fierz-identity, rewrite scalar products in terms of 
inverse propagators, and perform derivatives for $\hat\Pi_{ab}$ 
on the integrand level.
Using finally the on-shell condition $p^2=-m^2$, we obtain an intermediate 
result for the $\hat\Pi$ 
in terms of dimensionless one- and two-loop on-shell integrals $\hat I$,
defined as
\ba
\hat I(s_1,\dots,s_4) &\equiv&
\frac1{J(d,1)}
\int\frac{{\rm d}^dk}{(2\pi)^d} 
\ld \frac1{[k^2+s_3]^{s_1}}\,\frac1{[(k-p)^2+s_4]^{s_2}} \right|_{p^2=-1}\\
\hat I(s_1,\dots,s_{10}) &\equiv&
\frac1{\lk J(d,1)\rk^2}
\int\frac{{\rm d}^dk_1}{(2\pi)^d} 
\int\frac{{\rm d}^dk_2}{(2\pi)^d}\,
\frac1{[k_1^2+s_6]^{s_1}}\,
\frac1{[k_2^2+s_7]^{s_2}}\,
\times\nn&&\times
\frac1{[(k_1-k_2)^2+s_8]^{s_3}} 
\frac1{[(k_1-p)^2+s_9]^{s_4}} 
\ld 
\frac1{[(k_2-p)^2+s_{10}]^{s_5}} 
\right|_{p^2=-1} \;,
\ea
where the normalisation factor $J$ is a one-loop massive tadpole 
as defined in \eq\nr{eq:J}.

In a next step, using symmetries and reduction relations following
from systematic use \cite{tarasov,laporta} 
of integration-by-parts (IBP) identities \cite{ibp}, we
arrive at $d$\/-dimensional expressions in terms of a few master integrals,
as listed below.


\subsection{One-loop computations}

Applying the necessary projectors on the sum of diagrams depicted in 
\fig\ref{fig:1S0CT} as well as its $p^2$\/-derivative (which we take
at the integrand level)
and performing a reduction to master integrals, 
we obtain the $d$\/-dimensional 
results
\ba
\la{eq:PiT1}
\hat\Pi_{00}^{(1)} &=&
a_1\,K_1
+a_2\,K_2
\;,\\
\la{eq:PiTp1}
\hat\Pi_{10}^{(1)} &=&
b_1\,K_1
+b_2\,K_2
+b_3\,K_1^\prime
+b_4\,K_2^\prime
\;,
\ea
with master integrals $K_i$ listed in \app\ref{se:masters} and coefficients
\begin{align}
 a_1&= \frac98\,(4d-5)\;,
&a_2&= \frac14\,(2d-3)(2d-5)\;,\\ 
 b_1&= \frac3{16}\,(d-2)(4d-5)\;,
&b_2&= \frac18\,(12d^2-31d+18)\;,\\ 
 b_3&= \frac14\,(1-4\xi)\;,
&b_4&= \frac12\,(3-2d)\;.
\end{align}
Similarly, either from the diagrams of \fig\ref{fig:1S1CT} and using
\eq\nr{eq:Pi11CT} in reverse, or directly from the $m^2$\/-derivative
(easily taken at the integrand level)
of the first four diagrams of \fig\ref{fig:1S0CT}, 
\ba
\la{eq:PiT1p}
\hat\Pi_{01}^{(1)} &=&
4b_1\,K_1
+b_5\,K_2
+b_3\,K_1^\prime
+b_4\,K_2^\prime
\;,\\
b_5&=& \frac12\,(d^3-3d^2+4d-3)\;. 
\ea
Note that $\hat\Pi_{00}^{(1)}$ is gauge parameter independent, 
while $\hat\Pi_{10}^{(1)}$ and $\hat\Pi_{01}^{(1)}$ are not. 
However, their difference
$\hat\Pi_{10}^{(1)}-\hat\Pi_{01}^{(1)}=
-\frac{d-2}2\hat\Pi_{00}^{(1)}$
is gauge invariant. 
Let us remark that this relation between the three $\hat\Pi^{(1)}$ 
is not just a coincidence, 
but the one-loop on-shell case of a general relation which follows from
using that dimensional analysis gives $\PiT\sim{\rm mass}^2$
and $g^2\sim{\rm mass}^{4-d}$, such that
\be
\la{eq:derivs}
\(p^2\partial_{p^2}+m^2\partial_{m^2}+\frac{4-d}{2}\,g^2\partial_{g^2}\)
\PiT(p^2,m^2,g^2)=1\cdot\PiT(p^2,m^2,g^2) \;.
\ee


\subsection{Two-loop computations}

Applying the necessary projectors on the sum of diagrams depicted in 
\fig\ref{fig:2S0CT} and performing a reduction to master integrals, 
we obtain the $d$\/-dimensional result
\ba
\la{eq:PiT2}
\hat\Pi_{00}^{(2)} &=&
 c_1\,K_3
+c_2\,K_4
+c_3\,K_5
+c_4\,K_6
+c_5\,K_1\,K_1
+c_6\,K_1\,K_2
+c_7\,K_2\,K_2
\nn&&{}
-a_1\,b_3\,K_1\,K_1^\prime
-a_2\,b_3\,K_2\,K_1^\prime
-a_1\,b_4\,K_1\,K_2^\prime
-a_2\,b_4\,K_2\,K_2^\prime
\;,
\ea
with master integrals $K_i$ listed in \app\ref{se:masters} and coefficients
\ba
c_1&=& \frac3{64}\,(d-1)(176d-245)\;,\\ 
c_2&=& -\frac3{64}\,(144d^3-712d^2+1241d-760)\;,\\ 
c_3&=& -\frac{10800d^4-70632d^3+165227d^2-166654d+61752}{192(3d-4)}\;,\\ 
c_4&=& -\frac3{64}\,(d-2)(32d^3-312d^2+656d-405)\;,\\ 
c_5&=& \frac3{128}\,(32d^2-148d+155)\;,\\ 
c_6&=& -\frac3{16}\,(16d^4-188d^3+668d^2-940d+465)\;,\\ 
c_7&=& -\frac1{32}\,\frac{2d-3}{3d-4}\,(24d^5-164d^4+452d^3-680d^2+597d-242)
\;. 
\ea
Individual diagrams do have contributions to $\hat\Pi_{00}^{(2)}$
that are proportional to $1/N^4$ or $1/N^2$, 
but these cancel in the sum, leaving $\hat\Pi_{00}^{(2)}$ $N$\/-independent.

Note that $\hat\Pi_{00}^{(2)}+\hat\Pi_{00}^{(1)}\hat\Pi_{10}^{(1)}$ 
as well as $\hat\Pi_{00}^{(2)}+\hat\Pi_{00}^{(1)}\hat\Pi_{01}^{(1)}$ 
are gauge parameter independent.


\subsection{Results in 3d}

Let us here collect the expansions around $d=3-2\e$ of 
\eqs\nr{eq:PiT1}, \nr{eq:PiTp1} and \nr{eq:PiT1p}
\ba
\hat\Pi_{00}^{(1)}&\approxDDD&-\frac{63}{16}\ln3+\frac34+\order{\e}
\;\approx\;-3.57579+\order{\e}\\
\hat\Pi_{10}^{(1)}&\approxDDD&-\frac{21}{32}\ln3+\frac{33}8
-\frac{1-4\xi}8\ln\frac{2\sqrt\xi+1}{2\sqrt\xi-1}-\frac32\sqrt\xi
+\order{\e}\\
\hat\Pi_{01}^{(1)}&\approxDDD&-\frac{21}8\ln3+\frac92
-\frac{1-4\xi}8\ln\frac{2\sqrt\xi+1}{2\sqrt\xi-1}-\frac32\sqrt\xi
+\order{\e}
\ea
satisfying $\hat\Pi_{10}^{(1)}-\hat\Pi_{01}^{(1)}\;\approxDDD\;
-\frac12\hat\Pi_{00}^{(1)}+\order{\e}$,
as well as of \eq\nr{eq:PiT2} (subtracting $\xi$\/-dependence)
\ba
\la{eq:2loopExp}
\hat\Pi_{00}^{(2)}+\hat\Pi_{00}^{(1)}\hat\Pi_{10}^{(1)}&\approxDDD&
\frac3{20\e}
+\frac{849}{32}\frac{f(1/3)-f(7/9)}{\sqrt 2}
-\frac{1329}{512}\(-\frac{\pi^2}6+6\dilog(1/3)-2\dilog(1/9)\)
\nn&&
+\frac{17069}{4800}
+\frac{16761}{320}\ln2
-\frac{369}{8}\ln3
-\frac{9}{512}\ln^23
+\order{\e}\\
&\approx&\frac3{20\e}-10.6452+\order{\e}\;.
\ea
For comparison, $\hat\Pi_{10}^{(1)}=4\pi f_2(\xi)+\order{\e}$
in the notation of \eq(21) in \cite{eber}, 
but $\hat\Pi_{01}^{(1)}\neq-4\pi f_1(\xi)+\order{\e}$,
pointing to an error in that reference.


\section{Master integrals}
\la{se:masters}

We like to work with dimensionless and measure independent integrals, 
so let us normalise each loop by the massive one-loop tadpole integral 
$J(d,m)$, which for our choice of measure reads
\ba
\la{eq:J}
J(d,m)\;\equiv\; J 
&\equiv& \int\!\!\frac{{\rm d}^dk}{(2\pi)^d}\,\frac1{k^2+m^2}
= \frac1{m^2}\(\frac{m^2}{4\pi}\)^{d/2}\Gamma(1-d/2) 
\nn
&\approxDDD& -\frac{m}{4\pi}\(\frac{\pi e^{-\gamma}}{m^2}\)^\e
\(1+2\e+\order{\e^2}\)\;.
\ea

From reference \cite{raj}
(and using \eq(1) therein as well as \eqs(1,3,9-11,14) of 
\cite{broadhurst} for $K_3$), we get 
expansions around $d=3-2\e$ 
for all on-shell master integrals that we need:
\ba
K_1 
&=& \lk m^2 \TopoSB(\Lsc,\Asc,\Asc)/J \rk_{\rm os} 
= \hat I(1,1,1,1)\\ 
&\approxDDD& -\frac{\ln3}2  +\order{\e}\;,\\
K_1^\prime
&=& \lk m^2 \TopoSBlabel(\Lsc,\Asc,\Asc,{\xi m^2},{\xi m^2})/J \rk_{\rm os}
= \hat I(1,1,\xi,\xi) \\
&\approxDDD& -\frac12\ln\frac{2\sqrt{\xi}+1}{2\sqrt{\xi}-1} +\order{\e}\;,\\
K_1^{\prime\prime}
&=& \lk m^2 \TopoSBlabel(\Lsc,\Asc,\Asc,{},{\xi m^2})/J \rk_{\rm os}
= \hat I(1,1,1,\xi) \\
&\approxDDD& -\frac12\ln\frac{\sqrt{\xi}+2}{\sqrt{\xi}} +\order{\e}\;,\\
K_3 
&=& \lk m^6 \ToptSM(\Lsc,\Asc,\Asc,\Asc,\Asc,\Lsc)/J^2 \rk_{\rm os}
=\hat I(1,1,1,1,1,1,1,1,1,1)\\
&\approxDDD& \frac{f(1/3)-f(7/9)}{\sqrt{2}}  +\order{\e}
\quad{\rm with}\quad f(x) \equiv \Im\,\dilog(x+i\sqrt{1-x^2})\la{eq:fx}
\\&\approx& 
16\pi^2\times 0.000245310499\dots+\order{\e} \;,\\
K_4 
&=& \lk m^4 \ToptSAl(\Lsc,\Asc,\Asc,\Asc,\Asc)/J^2 \rk_{\rm os}
=\hat I(1,1,1,0,1,1,1,1,0,1)\\
&\approxDDD& \frac18\lk\ln^23-\frac{\pi^2}6+6\dilog(1/3)-2\dilog(1/9)\rk +\order{\e} \\
&\approx& 16\pi^2\times 0.00121156\dots  +\order{\e} \;,\\
K_5 
&=& \lk m^2 \ToptSS(\Lsc,\Asc,\Asc,\Lsc)/J^2 \rk_{\rm os}
=\hat I(1,0,1,0,1,1,0,1,0,1)\\
&\approxDDD& \frac1{4\e} +\(\frac12-2\ln2\)  +\order{\e}\;.
\ea

Two trivial one-loop massive vacuum master integrals read
\ba
K_2 
&=& \lk \TopoST(\Lsc,\Asc)/J \rk 
=\hat I(1,0,1,0) 
\;=\;1 \;,\\
K_2^\prime
&=& \lk \TopoSTlabel(\Lsc,\Asc,{\xi m^2}) /J\rk
=\hat I(1,0,\xi,0) 
\;=\;\xi^{(d-2)/2} \;.
\ea

One non-trivial fully massive vacuum master integral, 
expanded around $d=3-2\e$,
reads (cf. reference \cite{3dSunset}):
\ba
K_6 &=& \lk m^2\ToptVS(\Asc,\Asc,\Lsc)/J^2\rk = \hat I(1,1,1,0,0,1,1,1,0,0)\\
&\approxDDD& \frac1{4\e}-\(\frac12+\ln\frac32\)+\order{\e} 
\;\approx\; \frac1{4\e}-0.9054651081+\order{\e} \;.
\ea


\end{appendix}




\begin{thebibliography}{99}

\bibitem{linde}
  A.~D.~Linde,
  Phys.\ Lett.\  B {\bf 96} (1980) 289.

\bibitem{py}
  D.~J.~Gross, R.~D.~Pisarski and L.~G.~Yaffe,
  Rev.\ Mod.\ Phys.\  {\bf 53} (1981) 43.

\bibitem{teresa}
  A.~Cucchieri, D.~Dudal, T.~Mendes and N.~Vandersickel,
  arXiv:1202.0639 [hep-lat].

\bibitem{op01}
O.~Philipsen,
  Nucl.\ Phys.\ B {\bf 628} (2002) 167
  [hep-lat/0112047].

\bibitem{kara}
  D.~Karabali and V.~P.~Nair,
  Nucl.\ Phys.\ B {\bf 464} (1996) 135
  [hep-th/9510157].

\bibitem{Nair:2012zb}
  V.~P.~Nair,
  arXiv:1201.0977 [hep-th].

\bibitem{Kondo:2012wf}
  K.-I.~Kondo,
  arXiv:1202.4162 [hep-th].
    
\bibitem{bp}
  W.~Buchm\"uller and O.~Philipsen,
  Nucl.\ Phys.\  B {\bf 443} (1995) 47
  [arXiv:hep-ph/9411334];
  O.~Philipsen,
  hep-ph/9406307.

\bibitem{an}
  G.~Alexanian and V.~P.~Nair,
  Phys.\ Lett.\  B {\bf 352} (1995) 435
  [arXiv:hep-ph/9504256].

\bibitem{jp}
  R.~Jackiw and S.-Y.~Pi,
  Phys.\ Lett.\  B {\bf 403} (1997) 297-303
  [arXiv:hep-th/9703226].

\bibitem{corn}
  J.~M.~Cornwall,
  Phys.\ Rev.\  D {\bf 10} (1974) 500.

\bibitem{karsch}
  F.~Karsch, A.~Patkos and P.~Petreczky,
  Phys.\ Lett.\ B {\bf 401} (1997) 69
  [hep-ph/9702376].
  
\bibitem{andersen}
  J.~O.~Andersen and L.~Kyllingstad,
  Phys.\ Rev.\ D {\bf 78} (2008) 076008
  [arXiv:0805.4478 [hep-ph]].

\bibitem{blaizot}
  J.~P.~Blaizot, E.~Iancu and A.~Rebhan,
  Phys.\ Rev.\ D {\bf 68} (2003) 025011
  [hep-ph/0303045].

\bibitem{ja2}
  J.~O.~Andersen, M.~Strickland and N.~Su,
  arXiv:0911.0676 [hep-ph].

\bibitem{eber}
  F.~Eberlein,
  Phys.\ Lett.\ B {\bf 439} (1998) 130
  [hep-ph/9804460].

\bibitem{bir}
  J.~P.~Blaizot, E.~Iancu and A.~Rebhan,
  arXiv:hep-ph/0303185.
  
\bibitem{brs}
  C.~K.~Becchi, A.~Rouet and R.~Stora,
  Annals\ Phys.\ {\bf 98} (1976) 287-321.

\bibitem{zinnj}
  J.~Zinn-Justin,
  {\em Quantum field theory and critical phenomena},
  Int.\ Ser.\ Monogr.\ Phys.\  {\bf 113} (2002) 1.

\bibitem{rebhan}
  A.~K.~Rebhan,
  Phys.\ Rev.\ D {\bf 48} (1993) 3967
  [hep-ph/9308232].

\bibitem{jeger}
  F.~Jegerlehner, M.~Y.~Kalmykov and O.~Veretin,
  Nucl.\ Phys.\ B {\bf 641} (2002) 285
  [hep-ph/0105304].

\bibitem{Philipsen:2009dd}
  O.~Philipsen, D.~Bieletzki and Y.~Schr\"oder,
  PoS QCD {\bf -TNT09} (2009) 052
  [arXiv:0911.3595 [hep-ph]].
   
\bibitem{qgraf}
  P.~Nogueira,
  J.\ Comput.\ Phys.\  {\bf 105} (1993) 279;
  Nucl.\ Instrum.\ Meth.\ A {\bf 559} (2006) 220.

\bibitem{tarasov}
  O.~V.~Tarasov,
  Nucl.\ Phys.\ B {\bf 502} (1997) 455
  [hep-ph/9703319].

\bibitem{laporta}
  S.~Laporta,
  Int.\ J.\ Mod.\ Phys.\ A {\bf 15} (2000) 5087
  [hep-ph/0102033].

\bibitem{ibp}
  K.~G.~Chetyrkin and F.~V.~Tkachov,
  Nucl.\ Phys.\ B {\bf 192} (1981) 159;
  F.~V.~Tkachov,
  Phys.\ Lett.\ B {\bf 100} (1981) 65.

\bibitem{raj}
  A.~K.~Rajantie,
  Nucl.\ Phys.\ B {\bf 480} (1996) 729
  [Erratum-ibid.\ B {\bf 513} (1998) 761]
  [hep-ph/9606216].

\bibitem{broadhurst}
  D.~J.~Broadhurst,
  hep-th/9806174.

\bibitem{3dSunset}
  Y.~Schr\"oder and A.~Vuorinen,
  hep-ph/0311323;
  K.~Kajantie, M.~Laine, K.~Rummukainen and Y.~Schr\"oder,
  JHEP {\bf 0304} (2003) 036
  [hep-ph/0304048].


\end{thebibliography}
\end{document}